\documentclass{PoS}
\usepackage{wrapfig}
\usepackage[utf8]{inputenc} 
\usepackage[T1]{fontenc} 
\usepackage{amsmath,amssymb}
\usepackage{palatino} 
\usepackage{tikz}
\usepackage{graphicx} 
\usepackage{cite}
\usepackage{float}
\usepackage{slashed}
\usepackage{tikzscale}
\usetikzlibrary{calc,fit}
\usetikzlibrary{decorations.pathmorphing}
\usetikzlibrary{decorations.pathreplacing,decorations.markings}

\title{String breaking with 2+1 dynamical fermions using the stochastic
LapH method}

\ShortTitle{String breaking with 2+1 dynamical fermions using the stochastic
LapH method}

\author{\speaker{Vanessa Koch},$^{a,b}$ John Bulava,$^ c$ Ben H\"orz,$^ d$ Francesco Knechtli,$^ b$   Graham Moir,$^ e$ Colin Morningstar,$^ f$ Mike Peardon$^ a$  \\
\llap{$^a$}School of Mathematics, Trinity College Dublin, Dublin 2, Ireland\\
\llap{$^b$}Dept. of Physics, University of Wuppertal, Gaussstrasse 20, D-42119 Germany\\
\llap{$^c$}CP3-Origins, University of Southern Denmark, Campusvej 55, 5230 Odense M, Denmark\\
\llap{$^d$}PRISMA Cluster of Excellence and Institute for Nuclear Physics, Johannes Gutenberg-Universit\"at, 55099 Mainz, Germany\\
\llap{$^e$}Dept. of Mathematics, Hurstpierpoint College, College Lane, Hassocks, West Sussex, BN6 9JS, United Kingdom \\
\llap{$^f$}Dept. of Physics, Carnegie Mellon University, Pittsburgh, PA 15213, USA\\

E-mail: \email{kochv@maths.tcd.ie}, \email{bulava@cp3.sdu.dk}, \email{hoerz@uni-mainz.de}, \email{knechtli@physik.uni-wuppertal.de}, \email{ graham.moir@hppc.co.uk}, \email{colin\_morningstar@cmu.edu}, \email{mjp@maths.tcd.ie}}

\abstract{The static potential $V(r)$ between a static quark and a static antiquark separated by a distance r is defined as the energy of the ground state of the system. As a consequence of confinement, the energy between the quark-antiquark pair is contained inside a color flux tube, which will break due to pair creation as soon as the energy is high enough. String breaking is manifested as a quantum-mechanical mixing phenomenon between different states containing two infinitely heavy quarks. We investigate this phenomenon with $N_\mathrm{f}=2+1$ flavors of dynamical fermions in the stochastic LapH framework, using an ensemble of gauge configurations generated through the CLS effort. We observe the effect of the third sea-quark flavor, which results in a second mixing-phenomenon.}

\FullConference{The 36th Annual International Symposium on Lattice Field Theory - LATTICE2018\\
		22-28 July, 2018\\
		Michigan State University, East Lansing, Michigan, USA.}

\begin{document}

\section{Introduction}
String breaking, the transition of the static quark anti-quark string into a static-light meson-antimeson system, provides an intuitive example of a strong decay and is one of the defining characteristics of a confining gauge theory. 
Traditionally, in order to explore string breaking on the lattice, the Wilson loop was used as an observable. The breaking of the string should manifest itself by rendering the potential constant above a certain threshold, because it exhibits screening after the string is broken and saturates towards twice the static-light meson mass $2E_{m_B}$. But this phenomenon could not be observed in early simulations, even if the potential was calculated for distances bigger than the estimated string breaking distance, see for example \cite{Glassner, heller}. One problem is the small signal to noise ratio for distances beyond 1fm. However, the most important reason for the lack of evidence for string breaking using Wilson loops is that the Wilson line operator has a very small overlap with the ground state after the string is broken. 

In the theory with dynamical quarks, string breaking is manifested as a quantum-mechanical mixing phenomenon. This means that the two states, the string state as well as the two-meson state, are both needed to describe the potential. After the string is broken, the meson state dominates the new ground state of this system. In the neighborhood of the critical separation, the two states mix. If there is mixing, the ground state and first excited state are superpositions of the string state and the two-meson state, as a consequence the system undergoes an avoided level crossing, giving rise to an energy gap between the energy eigenstates. 

To examine the ground state and first excited state of the static potential as a mixing phenomenon, their energies are determined by a variational technique from a correlation matrix. Using this method, string breaking has been thoroughly investigated for the SU(2) Higgs model, e.g. \cite{knechtli1,knechtli2,philipsen}. For QCD with $N_\textrm{f}$=2 quark flavors, the most recent study of string breaking was performed on a single ensemble \cite{bali}, with sea quark mass slightly below the mass of the strange quark. So far, string breaking has not yet been observed for the N$_\textrm{f}$=2+1 theory. When the strange quark is included in the sea, two separate thresholds are expected, one for the decay into two static-light mesons and one for the decay into two static-strange mesons.
We investigate the phenomenon with N$_\textrm{f}$=2+1 flavors of non-perturbatively $O(a)$-improved dynamical Wilson fermions using an ensemble of gauge configurations generated through the Coordinated Lattice Simulations~\,(CLS) effort \cite{cls}. The N200 ensemble has a lattice size of $N_t \times N_s^3 = 128 \times 48^3$ with an estimated  isotropic lattice spacing of $a\approx 0.064$\,fm, pion mass $m_{\pi}=280$\,MeV and kaon mass $m_{K}=460$\,MeV. We employ the stochastic LapH method \cite{morningstar} in order to facilitate all-to-all propagation and calculate correlation functions required for string breaking efficiently. A variational analysis is used to extract the ground state as well as the first and second excited state of the system containing two static quarks. 

\section{String breaking as a mixing phenomenon}
Consider the potential of a system containing a heavy quark $Q(\mathbf{x},t)$ at point$(\mathbf{x},t)$ and a heavy anti-quark $\overline{Q}(\mathbf{y},t)$ at point $(\mathbf{y},t)$ in the static approximation. They are separated by a distance $r=\mid\mathbf{y}-\mathbf{x}\mid$, with conserved quantum numbers $\mathbf{x}$ and $\mathbf{y}$.
To examine the ground state of this system, i.e. the static potential, as well as the first and second excited state as a mixing phenomenon, the energies are determined by a variational technique from a correlation matrix. The matrix is built by correlation functions containing the interpolators for a Wilson loop, the two static-light and two static-strange meson state, where it is essential that the states have the same symmetries and thus carry the same quantum numbers. For every distance, a separate variational analysis has to be performed, so in order to extract the ground state as well as the first and second excited state a generalized eigenvalue problem (GEVP) \cite{gevp,luscher_gevp} is solved for each $r$.

If the interpolators $\mathcal{O}_W$, $\mathcal{O}_{B\overline{B}}$ and $\mathcal{O}_{B_s\overline{B}_s}$correspond to the string state and the state consisting of two static-light and two static-strange mesons respectively, the mixing matrix is given by
\begin{align}
C(\mathrm{r},t)=&\left(\begin{array}{ccc}C_{Q\overline{Q}}=\langle \mathcal{O}_{W}(t)\overline{\mathcal{O}}_{W}(0)\rangle
&C_{B\overline{Q}}=\langle\mathcal{O}_{B\overline{B}}(t)\overline{\mathcal{O}}_{W}(0)\rangle& C_{B_s\overline{Q}}=\langle\mathcal{O}_{B_s\overline{B}_s}(t)\overline{\mathcal{O}}_{W}(0)\rangle \\ \nonumber
C_{Q \overline{B}}=\langle \mathcal{O}_{W}(t)\overline{\mathcal{O}}_{B\overline{B}}(0)\rangle& C_{B\overline{B}}=\langle\mathcal{O}_{B\overline{B}}(t)\overline{\mathcal{O}}_{B\overline{B}}(0)\rangle& C_{B\overline{B}_s}=\langle\mathcal{O}_{B\overline{B}}(t)\overline{\mathcal{O}}_{B_s\overline{B}_s}(0)\rangle\\
C_{B\overline{Q}}=\langle\mathcal{O}_{B\overline{B}}(t)\overline{\mathcal{O}}_{W}(0)\rangle &C_{B_s\overline{B}}=\langle\mathcal{O}_{B_s\overline{B}_s}(t)\overline{\mathcal{O}}_{B\overline{B}}(0)\rangle&C_{B_s\overline{B}_s}=\langle\mathcal{O}_{B_s\overline{B}_s}(t)\overline{\mathcal{O}}_{B_s\overline{B}_s}(0)\rangle\end{array}\right)\\ 
= &\left(\begin{array}{lll}
\phantom{\sqrt{12}\times} \begin{tikzpicture}[baseline={([yshift=-.1ex]current bounding box.center)}]
\draw[gray, thick] (0,0) -- (0,1);
\draw[gray, thick] (0,1) -- (1,1);
\draw[gray, thick] (1,1) -- (1,0);
\draw[gray, thick] (1,0) -- (0,0);
\end{tikzpicture}
& \phantom{-}\sqrt{2}\times \begin{tikzpicture}[baseline={([yshift=-.8ex]current bounding box.center)}]
\draw[gray, thick] (0,0) -- (0,1);
\draw [decorate, decoration={snake, segment length=1.5mm, amplitude=0.3mm}](0,1) -- (1,1);
\draw[gray, thick] (1,1) -- (1,0);
\draw[gray, thick] (1,0) -- (0,0);
\end{tikzpicture} 
& \phantom{-\sqrt{21}\times} \begin{tikzpicture}[baseline={([yshift=-.8ex]current bounding box.center)}]
\draw[gray, thick] (0,0) -- (0,1);
\draw [decorate, decoration={snake, segment length=1.5mm, amplitude=0.8mm}](0,1) -- (1,1);
\draw[gray, thick] (1,1) -- (1,0);
\draw[gray, thick] (1,0) -- (0,0);
\end{tikzpicture} \\&\\
\phantom{'}\sqrt{2}\times \begin{tikzpicture}[baseline={([yshift=-.8ex]current bounding box.center)}]
\draw[gray, thick] (0,0) -- (0,1);
\draw[gray, thick] (0,1) -- (1,1);
\draw [gray, thick](1,1) -- (1,0);
\draw[decorate, decoration={snake, segment length=1.5mm, amplitude=0.3mm}](1,0) -- (0,0);
\end{tikzpicture}
& \phantom{--} 2 \times \begin{tikzpicture}[baseline={([yshift=-.8ex]current bounding box.center)}]
\draw [gray, thick](0,0) -- (0,1);
\draw [decorate, decoration={snake, segment length=1.5mm, amplitude=0.3mm}](0,1) -- (1,1);
\draw [gray, thick](1,1) -- (1,0);
\draw [decorate, decoration={snake, segment length=1.5mm, amplitude=0.3mm}](1,0) -- (0,0);
\end{tikzpicture}+\begin{tikzpicture}[baseline={([yshift=-.8ex]current bounding box.center)}]
\draw [gray, thick](0,0) -- (0,1);
\draw (0,0) .. controls (0.2,0.5) .. (0,1)[decorate, decoration={snake, segment length=1.5mm, amplitude=0.3mm}];
\draw [gray, thick](1,1) -- (1,0);
\draw (1,0) .. controls (0.8,0.5) .. (1,1)[decorate, decoration={snake, segment length=1.5mm, amplitude=0.3mm}];
\end{tikzpicture} 
& \phantom{12}\sqrt{2}\times \begin{tikzpicture}[baseline={([yshift=-.8ex]current bounding box.center)}]
\draw[gray, thick] (0,0) -- (0,1);
\draw [decorate, decoration={snake, segment length=1.5mm, amplitude=0.8mm}](0,1) -- (1,1);
\draw[gray, thick] (1,1) -- (1,0);
\draw[decorate, decoration={snake, segment length=1.5mm, amplitude=0.3mm}](1,0) -- (0,0);
\end{tikzpicture} \\&\\
\phantom{\sqrt{12}\times} \begin{tikzpicture}[baseline={([yshift=-.8ex]current bounding box.center)}]
\draw[gray, thick] (0,0) -- (0,1);
\draw[gray, thick] (0,1) -- (1,1);
\draw [gray, thick](1,1) -- (1,0);
\draw[decorate, decoration={snake, segment length=1.5mm, amplitude=0.8mm}](1,0) -- (0,0);
\end{tikzpicture}
& \phantom{-}\sqrt{2}\times \begin{tikzpicture}[baseline={([yshift=-.8ex]current bounding box.center)}]
\draw[gray, thick] (0,0) -- (0,1);
\draw [decorate, decoration={snake, segment length=1.5mm, amplitude=0.3mm}](0,1) -- (1,1);
\draw [gray, thick](1,1) -- (1,0);
\draw[decorate, decoration={snake, segment length=1.5mm, amplitude=0.8mm}](1,0) -- (0,0);
\end{tikzpicture}
& \phantom{-\sqrt{12}\times} \begin{tikzpicture}[baseline={([yshift=-.8ex]current bounding box.center)}]
\draw [gray, thick](0,0) -- (0,1);
\draw [decorate, decoration={snake, segment length=1.5mm, amplitude=0.8mm}](0,1) -- (1,1);
\draw [gray, thick](1,1) -- (1,0);
\draw [decorate, decoration={snake, segment length=1.5mm, amplitude=0.8mm}](1,0) -- (0,0);
\end{tikzpicture}+\begin{tikzpicture}[baseline={([yshift=-.8ex]current bounding box.center)}]
\draw [gray, thick](0,0) -- (0,1);
\draw (0,0) .. controls (0.2,0.5) .. (0,1)[decorate, decoration={snake, segment length=1.5mm, amplitude=0.8mm}];
\draw [gray, thick](1,1) -- (1,0);
\draw (1,0) .. controls (0.8,0.5) .. (1,1)[decorate, decoration={snake, segment length=1.5mm, amplitude=0.8mm}];
\end{tikzpicture}
\end{array}\right),\label{sbmat}
\end{align}

where the expectation value over gauge configurations is implicit in the pictorial representation of the diagrams.
The diagonal entries are given by the correlation functions for the string, the two static-light and static-strange meson state. The wiggly lines correspond to light and strange quark propagators, where the up and down quark are mass-degenerate sea quark flavors. Mixing occurs when the physical energy eigenstates are not unit vectors in this operator basis, which is shown explicitly by non vanishing off-diagonal elements.

\subsection{Steps towards string breaking}
The mixing matrix employed for the analysis is a $4 \times 4$ matrix. It is an extension of \ref{sbmat}, including 15 and 20 levels of spatial HYP smearing \cite{hyp} for $\mathcal{O}_{W}$. Following the method presented in \cite{wloops}, all gauge-links, including temporal links, are smeared using HYP2 parameters \cite{hyp2,staticdellamorte}: $\alpha_1 = 1.0, \alpha_2 = 1.0, \alpha_3 = 0.5$. This amounts to a change in the static action and a static propagator that is a modification of the static propagator derived by Eichten and Hill\cite{eichtenhill} to improve the signal-to-noise ratio at large Euclidean times \cite{staticdellamorte}. Afterwards, we construct a variational basis for the string state using 15 and 20 levels of HYP-smeared spatial links with parameters: $\alpha_2 = 0.6, \alpha_3 = 0.3$.

The diagrams involving light and strange quark propagators are measured on evenly spaced configurations of the N200 ensemble, employing a subset of the data used in \cite{christian}. Quark propagation is estimated using the stochastic LapH method \cite{morningstar} with parameters listed in table \ref{n200}, which are chosen in a way that they result in a similar physical
smearing as in previous studies \cite{morningstar}. The quark-field smearing makes use of stout-smeared gauge links \cite{stout} in order to spatially smooth the gauge field.
\begin{table}[H]
\centering
\small
\setlength{\tabcolsep}{.3pc}
\begin{tabular}{@{\extracolsep{0.0cm}}ccccccc}
\hline
id   &  $N_\mathrm{ev}$  &  $n_{\rho} \times  \rho$     & line type &  dilution scheme & $ N_r $ light/strange &   source time\\
 \hline
N200 &  192       &      $36 \times 0.1$       &  fixed    &  (TF,SF,LI8)	         &  5   /    2        &    32, 52     \\
     &            &                          &  relative &  (TI8,SF,LI8)              &  2   /    1        &        -
\end{tabular}
\caption{Number of eigenvectors, stout-smearing parameters, dilution
schemes, number of noise sources, and source times employed in this work. For definition of the LapH subspace and specification
of dilution schemes, see \cite{morningstar}.}\label{n200}
\end{table}
The GEVP is defined by 
\begin{equation}\label{fullgevp}
 C(t)\, v_n(t,t_0) = \lambda_n(t,t_0)\, C(t_0)\,v_n(t,t_0) \,, 
  \quad n=1,\ldots,N\,,\quad t>t_0 ,
\end{equation}
where $\lambda_n$ and $v_n$ are the eigenvalues and eigenvectors of the correlation matrix, respectively. For ${t\to\infty}$, the extracted eigenvalues are proportional to a single exponential function \cite{luscher_gevp}
\begin{equation}
 \lim_{t\rightarrow \infty}\lambda_n(t,t_0) = \exp(-E_n(t-t_0)). 
 \end{equation}
After solving the GEVP, three states are extracted using a two-parameter correlated fit to a single-exponential ansatz. Statistical uncertainties are estimated using bootstrap resampling \cite{bootstrap1,bootstrap2} with $N_{b}=800$ samples and the uncertainty quoted for numerical values is given by $1\sigma$  bootstrap errors. The covariance matrix was estimated on the original data and is kept \lq frozen\rq on all samples.

In the case of the string breaking spectrum, we are interested in the difference between the energies of the ground, first and second excited state and twice the energy of the static-light meson $2E_{m_B}$, which can be directly extracted from ratio fits.
If correlations between weakly-interacting two static-light meson and single static-light meson correlation functions are taken into account, this allows for a more precise extraction of the energy difference. We define the ratio
\begin{align}\label{ratio}
 R_n(t)= \frac{\lambda_n(t,t_0)}{C_B^2(t)},
\end{align}
where $C_B(t)$ is the correlation function of the static-light meson.

It was found in \cite{bali} that the limiting factor of the statistical resolution is the precision of the Wilson loop data, which a preliminary analysis of our data corroborated. In order to enhance the precision, the Wilson loops are determined on a much larger set of configurations and are rebinned such that the center of the bin is aligned with one of the configurations of the smaller set the other diagrams are measured on.

\begin{figure}[H]
\begin{center}
 \includegraphics[width=0.8\linewidth]{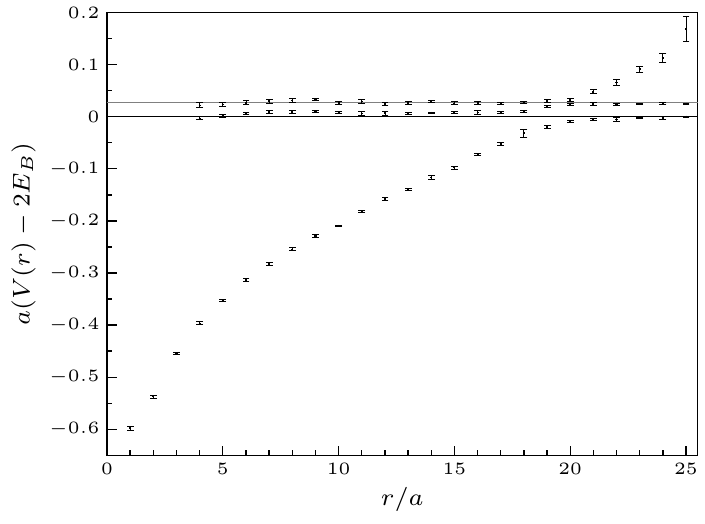}
 \caption{On-axis static potential determined using the full mixing matrix and renormalized by subtracting twice the energy of a static-light meson. The grey line corresponds to twice the static-strange mass, its error is too small to be visible. The error of the static-light meson mass is automatically taken into account by using the ratio given in equation (\ref{ratio}).}
\end{center}
\end{figure}
\begin{figure}[H]
\begin{center}
 \includegraphics[width=0.8\linewidth]{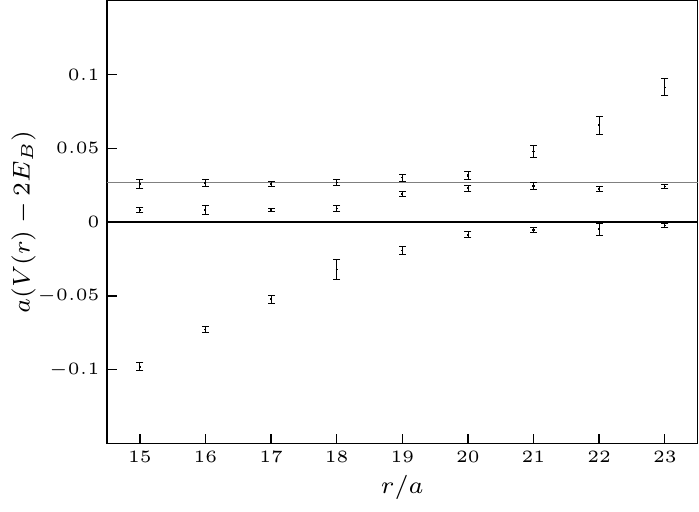}
\caption{Same as figure 1, zoomed into the string breaking region. }
\end{center}
\end{figure}
\section{Results and outlook}
Figure 1 and 2 show the preliminary results on 52 configurations for the diagrams involving light or strange quark propagators and 
Wilson loops measured on 1664 configurations, using only on-axis distances. The first and second excited state for the two smallest distances can not be reliably extracted, so there are no extracted energies shown for these states. For each distance $r$, we average over distinct permutations of $(0,0,r)$ in order to increase statistics by exploiting the cubic symmetry of the lattice. We do not observe any dependence on the direction in the data. 

The results of the preliminary mixing analysis are in agreement with the expected string breaking distance \cite{koch} given by the analysis of the potential using 100 configurations of N200, employing only rectangular Wilson-loops. The avoided level crossing between the ground state and the first excited state can be observed, whereas the expected second avoided crossing due to the formation of two static-strange mesons is not clearly visible using only on-axis distances. For distances beyond string breaking, the ground state tends towards the mass of two noninteracting static-light mesons, which cannot be observed if the potential is calculated from Wilson loops only. Apart from the string breaking region, the plot exhibits interesting behaviour for intermediate distances. 
The first excited energy differs from $2E_B$ for intermediate distances smaller than the string breaking distance, which could be an indication of interaction between the two static-light mesons. For the second extracted energy level, we observe agreement with $2E_{Bs}-2E_{B}$ at intermediate distances and beyond string breaking, as expected. 

The full mixing analysis with an enhanced spatial resolution as well as higher statistics is underway \cite{koch2}, utilizing a set of off-axis distances between $17a$ and $21.8a$ to ensure that we are able to map out the first as well as the second expected avoided level crossing.
\section*{Acknowledgements}
The authors wish to acknowledge the DJEI/DES/SFI/HEA Irish Centre for High-End Computing (ICHEC) for the provision of computational facilities and support. 
We are grateful to our colleagues within the CLS initiative for sharing ensembles. The code for the calculations using the stochastic LapH method is built on the USQCD QDP++/Chroma library \cite{chroma}. CJM acknowledges support by the U.S.~National Science Foundation under award PHY-1613449.
BH was supported by Science Foundation Ireland under Grant No.~11/RFP/PHY3218.  VK has received funding from the European Union's Horizon 2020 research and innovation programme under the Marie Sklodowska-Curie grant agreement Number 642069.

\bibliographystyle{JHEP}
\bibliography{sample}

\end{document}